# Adaptive Domain Model: Dealing With Multiple Attributes of Self-Managing Distributed Object Systems

Pavel Motuzenko, pavelm@softxp.com

Moscow State Institute of Electronic Technology

**Abstract.** Self-managing software has emerged as modern systems have become more complex. Some of the distributed object systems may contain thousands of objects deployed on tens or even hundreds hosts. Development and support of such systems often costs a lot. To solve this issue the systems, which are capable supporting multiple self-managing attributes, should be created. In the paper, the Adaptive domain concept is introduced as an extension to the basic domain concept to support a generic adaptation environment for building distributed object systems with multiple self-managing attributes.

## 1 Introduction

Modern distributed object systems may contain thousands of objects deployed on tens or even hundreds hosts. The increasing cost of managing the growing complexity of such systems is becoming an important issue influencing their future growth. Administration, support and maintenance of distributed systems often requires human intervention and costs a lot. To solve this issue, more autonomous systems should be created.

Today there are a number of solutions which already have the autonomic functions implemented. Among them are distributed multimedia applications which are capable adapting their behavior (and/or change their configuration) in accordance with the changing quality of network connections.

There are numerous application attributes of adaptation to changing runtime environment which should be taken into consideration by the application development team. Among them are: application self-configuring, self-optimization, automatic failure recovery, application self-healing [1]. In general, to deliver an application which is able to manage itself, a developer should learn how to deal with multiple adaptation attributes for an autonomous application with a structured approach.

The software logic for adapting to changing runtime environment may have different natures. For example, application failure recovery usually has a strategy allowing applications to react accordingly to node failures, process access violations, and so on. Whereas adaptation logic for software rejuvenation has a proactive strategy that detects the occurrence of software aging due to resource exhaustion, estimates the time remaining until the exhaustion reaches a critical point, and then automatically performs software rejuvenation procedures [2].

Complex distributed object systems may have a number of adaptation strategies concerning multiple attributes of system adaptation. Development of adaptation code for a distributed system can be subtle and error-prone as a developer does extra work implementing intricate multi-attribute adaptation logic across the system.

Our adaptive domain model depicts multiple attributes of distributed application adaptation into separate domains that represent a particular adaptation feature and implement adaptation logic for that feature.

## 2  Background and Related Works

### 2.1  Domains

The first specification representing a universal management architecture based on management domains was made by Becker K., Raabe U., M. Sloman and K. Twidle [3]. DOME (Domain-Oriented Management Environment) implements this architecture, and management domains are used for partitioning, structuring and naming managed objects. DOME supports policies for access, and domains are used to state the application of management policies. Managed objects in a domain do not share characteristics such as type, location or ownership, but are instead explicitly included into domains for the purposes of management.

Domain containment is implemented by domains holding a reference to each of their members, and does not represent physical containment. Domains give a local name for each member, unique within the individual domain naming context. All domains are reachable from a well-known root domain, path names representing traversals of the domain hierarchy can thus be used to refer uniquely to a managed object. Domains are managed objects in their own right, and subject to frequent, management via a domain service.

Halldor Fosså proposed extensions to the basic domain concept to support generic interactive management environment [4]. These include the capability for storing and retrieving type information for invocation support and graphical display, either from a managed object or from its parent domain. Enhancements to the architecture of the domain service were made, and an extended role of the per-host manager was introduced. Per-host manager provides support for detecting configuration failures. In this approach the domain structure is visible to managers in management applications via a user interface. Different management applications can represent domain structures in different ways, and different user interfaces are possible, including textual, graphical and three-dimensional interfaces.

### 2.2  Adaptable Middleware

dynamicTAO provides a component model to support the reconfiguration of the ORB [5]. The component model employs explicit representations of inter-component dependencies and component requirements. The representation can be manipulated in order to reconfigure the implementation of the ORB. Adaptation logic can be implemented by using several components which can be dynamically reconfigured to create new types of adaptation strategies. Adaptation logic code can be implemented in some centralized manner. There is no explicit support for adaptation logic with multiple self-managing attributes covering different sets of components.

K-Components is a component model for building context-adaptive applications [6]. Components are specified using a subset of IDL-3, called K-IDL, while their software architecture's adaptive behavior is specified in a separate Adaptation Contract Description Language (ACDL). Adaptation logic, encapsulated in the ACDL, can be written by programmers to build self-adaptive systems, but can also be modified and updated at runtime by users, allowing them overall control of the application's adaptive behavior. Systems can be adapted across many different layers of abstraction, including at the application level, system level and resources level. Adaptation policies are encapsulated as contracts between the different entities involved.

OpenCOM is a lightweight efficient component model based on the in-process implementation of Microsoft COM [7]. The component model also maintains explicit representations of inter-component dependencies. The representation can then be manipulated in an efficient manner. OpenCOM supports introspection, intercession and adaptation capabilities by manipulating its fundamental terms: interfaces, connections and receptacles. OpenCOM has flexible reconfiguration support allowing third party dynamic reconfiguration of both components and their connections.

## 3 Adaptive domains

The adaptive domain concept is introduced as an extension to the basic domain concept to support a generic adaptation environment for building distributed object systems with multiple self-managing attributes.

Features extending basic domain concept [3] include support for separate loadable adaptation logic for each management domain, dynamic managed object inclusion/exclusion into domain, persistent domain configurations.

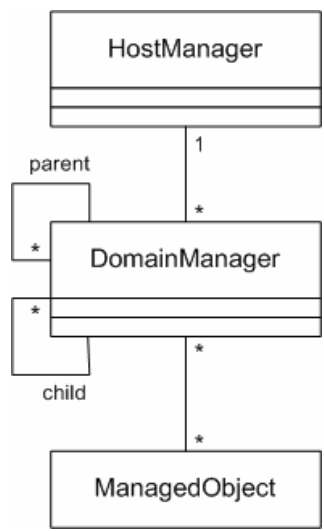

**Fig. 1. Conceptual view: domain manager**

Separate adaptation logic of a domain implements a particular self-managing feature of distributed object system. Adaptation logic for a domain resides in the domain manager and is made up of five main parts which collaborate to support decision-making: monitoring, auditing, analyzing, regulating, and adaptation execution code. All the code of adaptation logic is expressed in terms of domain direct and indirect member managed objects (using their path names), and operations on them. Adaptation logic can have the following adaptability strategies:

- Reactive, event-driven. Upon receiving the adaptation events, the system accordingly triggers its behavior or configuration at once.

- Proactive, preventive. Changes in the runtime environment are expected beforehand. A priori measurements are taken to adapt to the changes (e.g. a priori design).

- Retroactive. When changes take place in the runtime environment, the system does not respond to the adaptation events (new or modified), which is recognized; the system changes its behavior or configuration to meet the changed needs.

Domain adaptation logic consumes adaptation events from its internal and external sensors. One sensor may serve one or more domains. A sensor is a managed object itself so it may have several path names for each of the domains it belongs to. There are no restrictions for an ordinary managed object to act as a sensor emitting adaptation events for its parent domains.

Domain adaptation logic makes adaptation decisions, which are realized using actuators. We define three types of actuators: configuration manager, adaptation commands, and mobile adaptation agents.

The configuration manager implements dynamic reconfiguration in accordance with the system meta-model (e.g. "components and connections" graph) [6, 8]. Since in most component object frameworks such configuration graph can be built at runtime (e.g. when class name for an object is defined by the user), the configuration manager also deals with actual connections made. The configuration manager is a singleton object for a system executing one command at a time. That means that only one reconfiguration is possible at the same time, although there is a sufficient condition for concurrent graph reconfiguration (based on the blocking approach [9]) which allows concurrent execution of several reconfigurations.

Adaptation commands and mobile adaptation agents allow domains to propagate necessary changes to the managed objects structure. Adaptation commands are mainly used for communication between the parent domain and its child domains. Mobile adaptation agents are used for executing complicated, time-consuming changes within managed objects structures. An actuator is a managed object itself too so it may have several path names for each of the domains it belongs to. There are no restrictions for an ordinary managed object to act as an actuator executing adaptation changes according to its semantics.

Domain adaptation logic contains the monitoring and auditing code. The monitoring code of adaptation logic implements mechanisms that receive and collect adaptation events from sensors for the domain. The auditing code also implements sporadic audit functions on managed objects and sensors (sanity checks).

Domain adaptation logic contains the analyzing code. Analyzing code may implement adaptation-specific retrospective analysis, forecasting or model checks. Analyzing part of adaptation logic is responsible for delivering a decision for domain adaptation and its details. Analyzing part is also responsible for consistency preserving checks for a proposed decision.

Domain adaptation logic may contain the regulating code. Regulating code is responsible for representing adaptation scenarios based on adaptability strategies. For example, for proactive or retroactive adaptability strategy it may produce anti-oscillation action scenario for executing adaptation smoothly.

Domain adaptation logic always contains the adaptation execution code. Adaptation execution code is responsible for controlling the execution of adaptation by means of actuators according to the adaptation scenario.

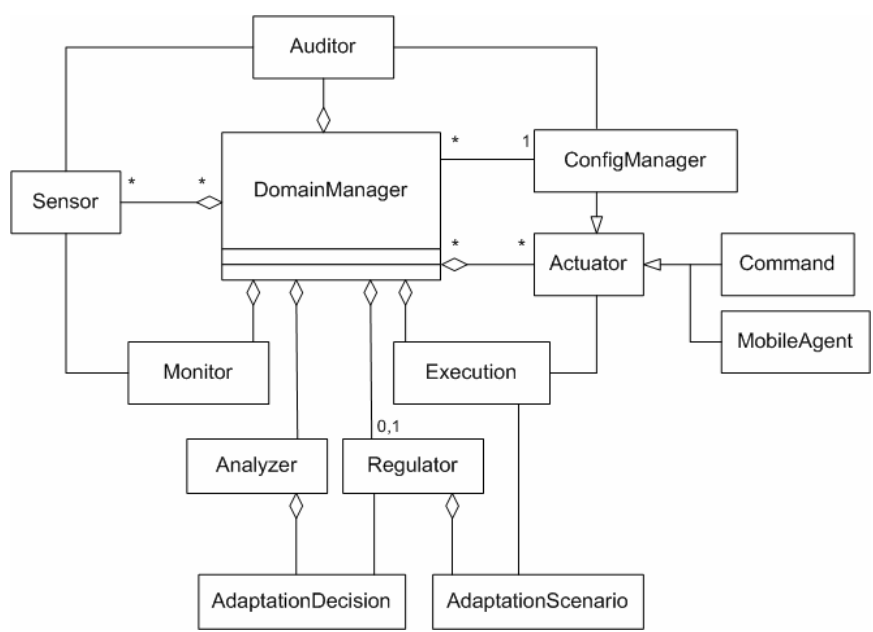

**Fig. 2. Conceptual view: adaptive domain design**

The adaptive domain is subject to adaptation management policies applied to it (i.e. explicit system adaptation). Adaptation management policies may contain high-level information specified by a human manager as well as low-level information specified by the parent domain to its child domain. Adaptive domains collaborate with each other using adaptation events (which can be propagated by a child to the parent domain), or adaptation commands (which can be generated by a parent domain to its child domains).

The code of the domain operates on path names of managed objects. In a simple case, that is the unique local name of the object within the domain scope. Due to internal domain services allowing for enumeration and browsing of objects with direct and indirect memberships, adaptation logic of the domain can be loosely-coupled to concrete object implementation. The loadable/unloadable code of adaptation logic thus can be independent of the specific managed objects structure allowing the developer to collect and reuse adaptation logic implemented once.

Dynamic managed object inclusion/exclusion into domain results in changes of the adaptation logic for that managed object. A managed object that was included in a domain obtains a new self-managing attribute, which wasn't available for that object before. In certain cases such an adaptation will work even if it wasn't designed by the developer for that managed object earlier.

Our Adaptive Domain Framework (ADF) prototype utilizes .NET Remoting. Its implementation is in C#. C# is also used for the development of adaptation logic of a framework domain set. The blocking transaction approach [9] is used for implementing reconfiguration, including concurrent system reconfiguration of multiple domains.

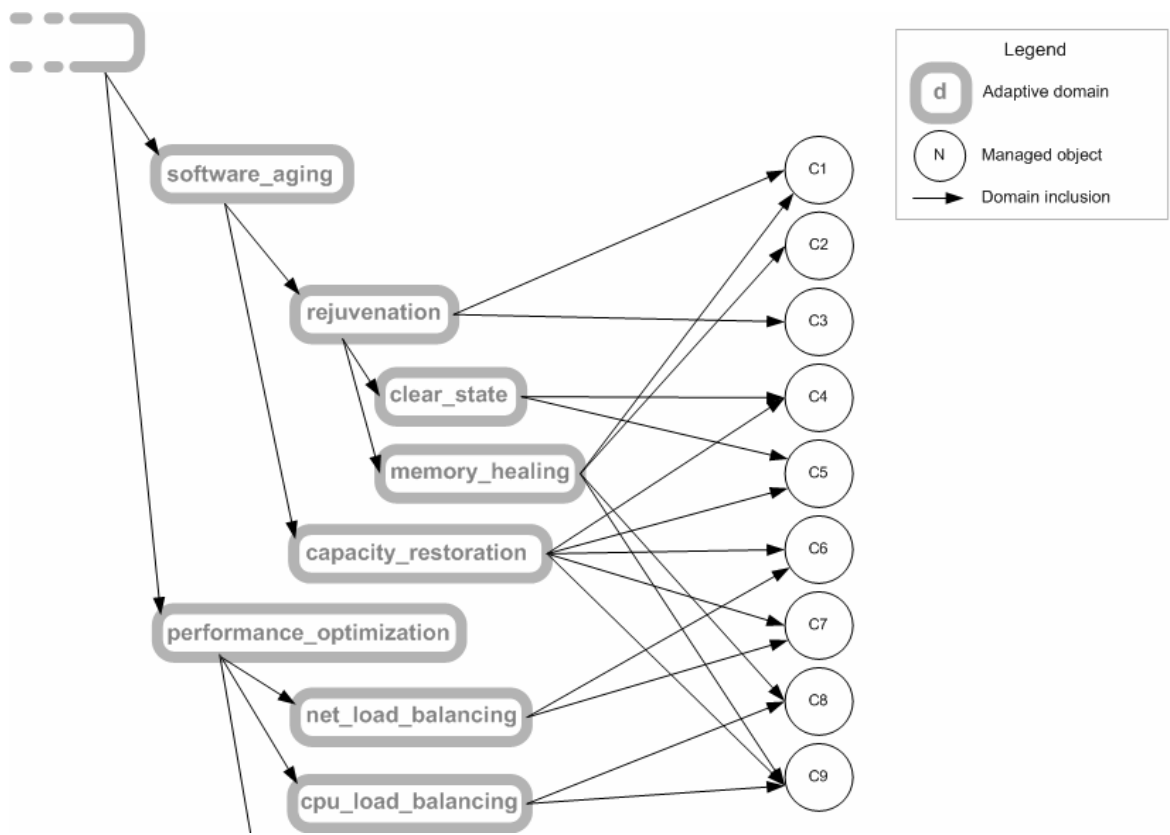

**Fig. 3.  Example of a domain hierarchy decomposing multiple self-managing attributes**

# 4 Conclusions

Without a mechanism for decomposing multiple self-managing attributes of distributed object systems, adaptation code specific for one attribute becomes tangled with plenty of adaptation code for other adaptations making development of a full-fledged self-managing application subtle and error-prone.

The adaptive domain concept is introduced as an extension to the basic domain concept to support a generic adaptation environment for building distributed object systems with multiple self-managing attributes. Our adaptive domain model depicts multiple attributes of distributed application adaptation into separate domains that represent a particular adaptation feature and implement adaptation logic for that feature.